\documentclass[a4paper,fleqn]{cas-sc}

\usepackage[numbers]{natbib}

\usepackage{amssymb}
\usepackage{amsmath}
\usepackage{tabularx}

\usepackage[table]{xcolor}
\definecolor{rowgray}{gray}{0.93}

\usepackage{array}
\usepackage{hyperref}
\usepackage{booktabs}
\usepackage{multirow}
\usepackage{graphicx}
\usepackage{float}

\hypersetup{
  colorlinks=true,
  linkcolor=blue!50!black,
  citecolor=green!40!black,
  urlcolor=blue!60!black
}
\usepackage{enumitem}
\setlist[itemize]{topsep=3pt,itemsep=2pt,parsep=0pt,leftmargin=1.5em}
\setlist[enumerate]{topsep=3pt,itemsep=2pt,parsep=0pt,leftmargin=1.8em}

\usepackage{float}
\makeatletter
\@ifpackageloaded{float}{%
  \restylefloat{figure}%
  \restylefloat{table}%
}{}
\makeatother
\begin{document}
\let\WriteBookmarks\relax

\shorttitle{Security and Safety Threats in Generative AI}

\title[mode=title]{From AI-Generated Content to Agentic Action: Security and Safety Threats in Generative AI}

\author[1]{Zelin Zhang}

\fnmark[1]
\ead{zelin.zhang@queensu.ca}
\ead[url]{}
\credit{}

\author[2]{Qi Li}

\fnmark[2]
\ead{qi.li@queensu.ca}
\ead[url]{}
\credit{}

\author[1]{Jie Cao}

\fnmark[3]
\ead{jie.cao@queensu.ca}
\ead[url]{}
\credit{}

\author[3]{Lingshuang Liu}

\fnmark[4]
\ead{lingshuang.liu@uwaterloo.ca}
\ead[url]{}
\credit{}

\author[1]{Jianbing Ni}
\cormark[1]
\fnmark[5]
\ead{Jianbing.ni@queensu.ca}
\ead[url]{}
\credit{}
\corref{cor}
\cortext[cor]{Corresponding author}

\affiliation[1]{organization={Department of Electricla and Computer Engineering, Queen's University},
            addressline={99 University Ave},
            city={Kingston},
            postcode={K7L 3N6},
            state={ON},
            country={Canada}}

 \affiliation[2]{organization={School of Computing, Queen's University},
            addressline={99 University Ave},
            city={Kingston},
            postcode={K7L 3N6},
         state={ON},
            country={Canada}}

\affiliation[3]{organization={Department of Electricla and Computer Engineering, University of Waterloo},
            addressline={200 University Ave West},
            city={Waterloo},
            postcode={N2L 3G1},
            state={ON},
            country={Canada}}

\begin{abstract}
Generative AI systems are increasingly used not only to produce
content but also to retrieve data, invoke tools, and execute actions.
This work examines the security and safety implications of that
shift across content-level, model-level, and agentic threats. We analyze how attacker access requirements, system autonomy, and the scope of potential harm change as models move from generating artifacts to executing operations through tool chains and external APIs. We then assess technical countermeasures including detection,
watermarking, alignment, and emerging agentic safeguards, and show
that several depend on forms of institutional coordination that
current governance arrangements do not yet provide. Across the cases
examined, capability deployment and attack-surface expansion
repeatedly outpace defensive responses as systems move from generating content to executing
real-world actions.
\end{abstract}


\maketitle

\section{Introduction}
\label{sec:introduction}

Generative AI (GenAI) has advanced rapidly in recent years, transforming digital content production from a labor-intensive human process into an increasingly automated and lower-cost one across text, images, audio, video, and code, often with quality that makes synthetic content difficult for humans to distinguish from authentic material. Nevertheless, the same capabilities that drive these advances have given rise to a new class of security and safety challenges. First, the unprecedented realism and accessibility of AI-generated content have enabled its large-scale exploitation for malicious purposes, including the fabrication of disinformation, impersonation, and IP theft. Second, AI systems themselves have emerged as high-value targets for adversarial manipulation, with documented vulnerabilities spanning training-time poisoning, inference-time evasion, and deployment-time abuse. Third, and perhaps most consequentially, the delegation of consequential real-world actions to autonomous AI agents introduces an operational risk frontier that extends far beyond the scope of prior content-generation paradigms. Together, these three dimensions, content-level weaponization, system-level adversarial exploitation, and agent-level operational risk, delineate the threat landscape that this paper systematically analyzes.

The consequences of this transformation are already visible in both model capability and threat potential.
Human observers identify AI-generated images correctly only 62\% of the time \cite{roca2025good}, underscoring how synthetic media now frequently evades unaided human judgment.
At the same time, the near-zero marginal cost of generation has enabled the mass production of malicious content, including phishing and social-engineering campaigns at a scale previously unattainable through purely human labor.
This shift represents a fundamental operational extension of generative AI which means in modern agentic systems, model outputs increasingly function not only as content but also as intermediate representations for planning, tool use, API interaction, and action execution.

In this paper, we extend the analytical scope of AI-Generated Content (AIGC) beyond synthetic artifacts in text, images, audio, video, and code to also include action, namely sequences of real-world operations carried out by autonomous AI agents through tool invocation, API requests, and shell commands. This framing reflects a structural shift in how generative systems are deployed, as LLM-based agent frameworks and vision-language-action models increasingly blur the boundary between generating a response and executing an operation. Additionally, the extension is analytical rather than terminological; we do not propose to redefine AIGC but use the content-to-action continuum as an organizing lens for threats that span both domains.

However, the limitations of existing work become clear.
Deepfake surveys, for example, have concentrated mainly on visual synthesis and detection \cite{pei2024deepfake}; they are far less informative about text-centered threats or autonomous agents.
Work on AI-enabled influence operations has been more attentive to disinformation \cite{spitale2023ai}, yet says little about model-level attacks or the technical defenses that respond to them.
Nor is the gap filled by widely cited practitioner resources. The OWASP Top~10 for LLM Applications \cite{owasp_llm_top10_2025} is invaluable as an operational checklist, but it does not aim to provide a scholarly taxonomy. The International AI Safety Reports \cite{bengio2025international,bengio2026international}, by contrast, offer broad and authoritative assessments of risk while stopping short of a sustained treatment of technical countermeasures. What is still missing is an end-to-end account that connects content-generation risks, adversarial attacks on models, and agentic exploitation, while also placing technical defenses in conversation with governance and ethics.

This work addresses that gap by offering an analysis of AIGC security and safety across content-level, model-level, and agentic threats, while bridging technical, governance, and ethical perspectives.
We make four contributions:
\begin{figure}[!htbp]
\centering
\includegraphics[width=0.95\linewidth]{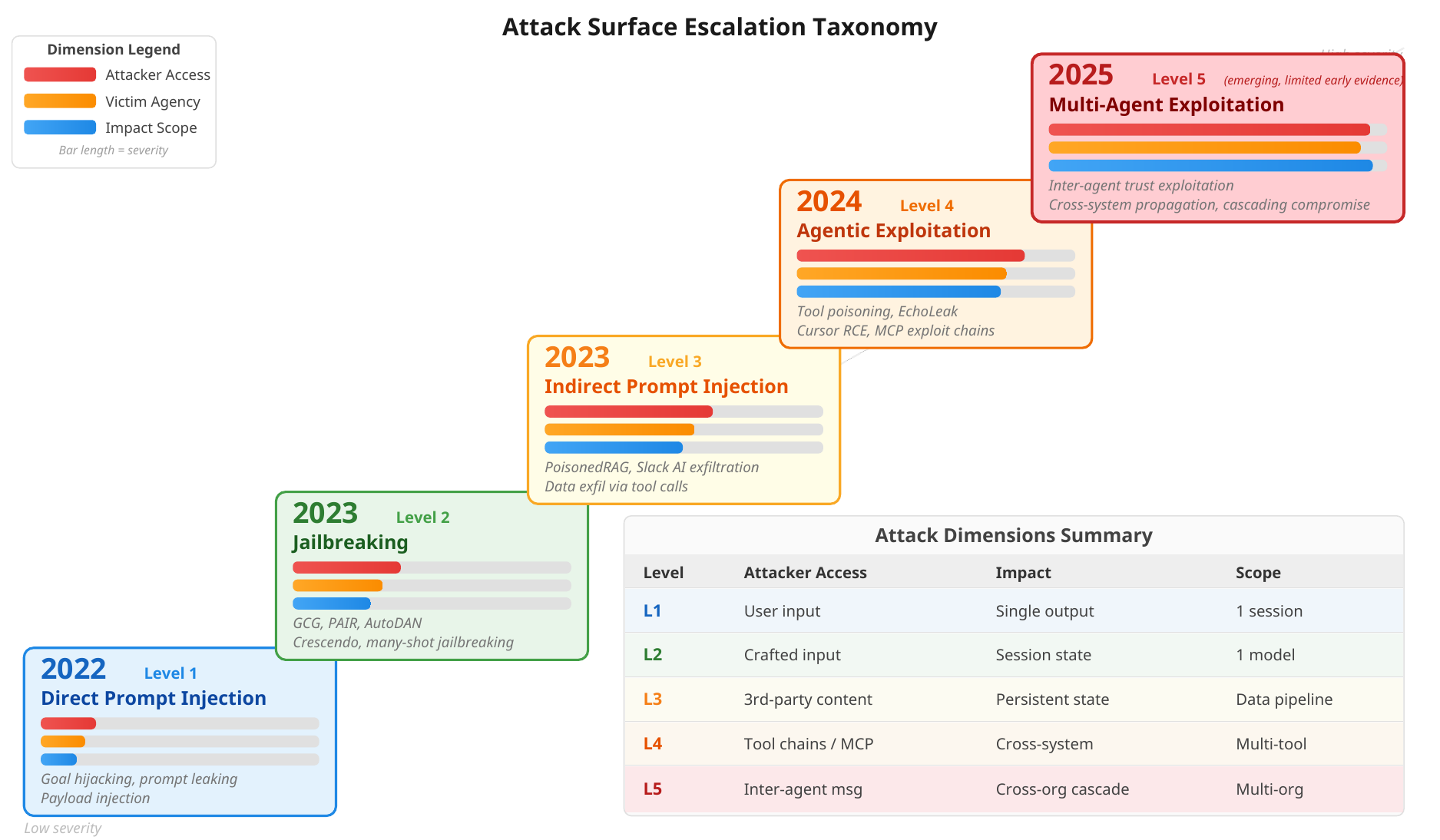}
\caption{Adversarial attacks organized by attacker access, victim agency, and impact scope, from direct prompt injection (2022) through jailbreaking, indirect prompt injection, and agentic exploitation to multi-agent attacks (2025--). Year labels indicate representative first-appearance dates, not a required attack sequence; these levels are partly orthogonal and may co-occur in deployment.}
\label{fig:evolution-ladder}
\end{figure}

\begin{enumerate}[nosep,leftmargin=1.5em]
    \item \textbf{A security–safety framing of AIGC threats and responses.}
    We use the distinction between security failures and safety harms to highlight a difference in response speed and response type. Security threats more often lead to technical countermeasures, while safety harms depend more on platform policy, cross-institutional coordination, and legislation. Governance delays therefore tend to weigh more heavily on the safety side. (Section~\ref{sec:threats}).

    \item \textbf{A cross-layer analytical synthesis of AIGC risk.}
   We place content-level weaponization, model-level attacks, and agentic exploitation within a single analytical frame organized around the shift from content generation to tool use and action execution. This framing emphasizes how changes in deployment architecture expand attacker access, increase system agency, and widen the downstream scope of harm across otherwise separate threat literatures (Sections~\ref{sec:cap-autonomy} and~\ref{sec:threat-adversarial}).

    \item \textbf{An integrated analysis of technical and governance responses.}
   By reading technical countermeasures alongside governance developments, we show that several promising defenses, including provenance mechanisms, watermark verification, and emerging MCP safeguards, presuppose forms of institutional coordination that current governance arrangements do not yet provide. The governance gap is therefore structural as well as temporal (Section~\ref{sec:countermeasures}, Section~\ref{sec:governance}).

     \item \textbf{Externalized model use as a change in security assumptions.}
    We argue that prompt injection, indirect prompt injection, tool poisoning, and related agentic attacks should be understood not simply as additional attack examples, but as consequences of a broader shift in how generative models are deployed. Once models access external data, invoke tools, and execute actions, defenses designed for input--output systems no longer transfer cleanly to connected and action-bearing environments (cross-cutting; especially Sections~\ref{sec:threat-adversarial} and~\ref{sec:future-agents}).
\end{enumerate}

This paper is an analytical review rather than a systematic survey. Its purpose is not exhaustive coverage of every subfield in generative AI security and safety, but a synthetic analysis of how the shift from content generation to tool use and agentic action reshapes the threat surface and the assumptions behind defense and governance. Source selection therefore follows a principle of representativeness rather than exhaustiveness. The evidence base includes peer-reviewed studies, institutional risk reports, vendor security disclosures, standards and specification documents, and documented incidents. These source types do not serve identical functions in the analysis. For fast-moving topics such as MCP, agent security incidents, protocol evolution, and governance updates, primary sources are used where they provide the most authoritative record of chronology, technical detail, or official policy. For broader analytical claims, however, peer-reviewed research and cross-source synthesis are given priority. In this sense, the paper complements rather than duplicates modality-specific surveys, practitioner checklists, or broad risk reports by bringing these partly separate discussions into a single analytical frame. Non-peer-reviewed sources are drawn primarily from the organizations that develop the systems and standards under discussion---including Anthropic, OpenAI, Google, Microsoft, NIST, OWASP, and the C2PA coalition---and are cited where they provide the authoritative record of protocol design, vulnerability disclosure, or policy development.

\section{Generative AI Capabilities and Threat-Relevant Properties}
\label{sec:capabilities}

This section identifies four properties of modern generative systems that convert rapid capability gains into concrete security risks, including high-fidelity generation (Section~\ref{sec:cap-fidelity}), low-cost scalable production (Section~\ref{sec:cap-scale}), fine-grained controllability (Section~\ref{sec:cap-control}), and autonomy with tool use (Section~\ref{sec:cap-autonomy}). The first three properties concern AI as a content producer, determining what can be generated, at what cost, and with what precision; the fourth marks the pivot to AI as an autonomous actor, where the threat shifts from harmful output to harmful action. Table~\ref{tab:cap-threat} maps each property to its primary threat vector and the section in which that threat is analyzed.

\begin{table}[!htbp]
\centering
\caption{Mapping generative AI capability components to representative security threats.}
\label{tab:cap-threat}

\fontsize{7}{7}\selectfont
\setlength{\tabcolsep}{5pt}
\renewcommand{\arraystretch}{1.3}

\begin{tabular}{p{3.8cm} p{4.8cm} p{4.5cm} c}
\toprule
\textbf{Capability Component} & \textbf{Key Metric} & \textbf{Primary Threat} & \textbf{\S} \\
\midrule

High-fidelity generation
& Human detection near chance (62\%)
& Synthetic-content deception
& \ref{sec:threat-deepfake} \\

Low-cost production
& $<$\$1 / M tokens (frontier mini models)
& Scalable phishing \& fraud
& \ref{sec:threat-phishing} \\

Controllability
& Few-shot identity adaptation
& Targeted impersonation
& \ref{sec:threat-ncii} \\

\midrule

Reasoning \& long context
& Frontier reasoning + 1M+ context
& Multi-step attack planning
& \ref{sec:threat-model} \\

Autonomous agency
& File / shell / GUI / API access
& Environment compromise
& \ref{sec:threat-ipi} \\

Tool ecosystem (MCP)
& 10{,}000+ public servers
& Tool poisoning \& cross-server attacks
& \ref{sec:threat-adversarial} \\

\bottomrule
\end{tabular}
\end{table}
\subsection{High-Fidelity Multi-Modal Generation}
\label{sec:cap-fidelity}

Generation fidelity across text, image, audio, and video has reached a point where unaided human judgment is no longer a reliable way to distinguish authentic from synthetic content, as controlled experiments consistently show detection accuracy at or near chance levels, and the modalities most central to identity verification, especially voice and video, can now be reproduced from only minimal reference material. The following paragraphs examine these two dimensions in turn, first by considering the collapse of perceptual discrimination in text and image and then by addressing the erosion of identity-based authentication through synthetic audio and video.

In the text modality, frontier LLMs produce paragraphs which are consistently difficult to distinguish from human output, which is close to chance. As a result, text-based social engineering such as phishing emails and fabricated legal documents can no longer be reliably identified by human reviewers alone. Synthetic images present a parallel challenge, as a 2025 Microsoft study found that participants correctly identified AI-generated images only 62\% of the time~\cite{roca2025good}, which suggests that latent diffusion and Diffusion Transformer architectures have already pushed image synthesis beyond the point at which human perception can reliably distinguish real from synthetic content. Automated detectors fare little better because robustness to common post-processing operations remains a critical and largely unsolved weakness~\cite{bengio2025international}. The problem also extends beyond detection rates, since decades of anti-phishing training taught employees to identify threats through spelling errors, generic greetings, and awkward phrasing, yet these cues have largely lost their value now that LLMs can produce grammatically polished and contextually appropriate text while diffusion models generate images without the artifacts that were common in earlier GAN-based systems.

In addition to making synthetic content difficult to detect, audio and video synthesis also undermine identity-based authentication. Neural codec language models such as VALL-E have demonstrated zero-shot voice cloning from as little as three seconds of reference audio~\cite{wang2023neural}, and earlier work has shown that intelligible speech can even be reconstructed from speaker embeddings alone, further undermining the assumed unidirectionality of voiceprint-based authentication~\cite{lu2021voxstructor}. Subsequent systems have approached human parity on standard speech-quality benchmarks, yet current voice verification systems still face error rates of up to 23\% when tested against high-quality synthetic speech~\cite{bengio2025international}, which weakens a channel that many organizations continue to treat as a trusted second factor. Video generators such as Sora~\cite{brooks2024video} produce temporally consistent and photorealistic output, compounding the erosion of the multi-layered trust infrastructure on which organizational security has historically depended. Across all four modalities, the reliability of human judgment, automated filters, and voice- and video-based authentication continues to decline—specific incidents illustrating the resulting harms are examined in Section~\ref{sec:threat-misinfo}.

Across all four modalities, the consistent finding is that generation fidelity has outpaced human detection capacity, removing a perceptual barrier that historically constrained the effectiveness of synthetic-content attacks. This asymmetry alone would pose a manageable challenge if producing high-fidelity content still required specialized expertise or substantial resources; as Section~\ref{sec:cap-scale} shows, neither condition holds.

\subsection{Low-Cost Scalable Production}
\label{sec:cap-scale}

The resource barrier has collapsed across three dimensions, namely access to open-weight models, consumer-grade local deployment, and hosted API inference, which together place state-of-the-art generative capabilities within reach of almost any motivated individual. This diffusion has been further accelerated by rapid improvements in the model infrastructure stack, including parameter-efficient fine-tuning, quantization, optimized inference kernels, attention acceleration, and KV-cache engineering, all of which reduce the hardware, memory, and latency requirements for running and adapting capable models. Each dimension lowers a different component of cost, and together they remove much of the economic and logistical friction that once limited advanced content generation to well-resourced actors.

Meta's Llama family helped normalize open-weight release at scale~\cite{grattafiori2024llama}, and Meta reported in March 2025 that Llama had surpassed one billion downloads~\cite{Meta2025LlamaBillion}. Since then, however, the Qwen family has become an equally important example of frontier capability diffusion in the open-weight ecosystem, with Qwen2.5 and Qwen3~\cite{hui2024qwen2,yang2025qwen3} spanning model sizes from sub-billion local models to frontier-scale mixtures of experts. This spread in model size matters because it brings capable systems within reach of both high-end local users and small-scale operators.

Once downloaded, open-weight models can be modified with little external oversight. Safety alignment and refusal behavior can be weakened or removed through relatively lightweight fine-tuning, producing uncensored variants that circulate in underground communities~\cite{bengio2025international}. Open weights therefore remove not only cost but also policy, because the refusal rules, terms of service, and usage logging enforced by cloud providers do not carry over to local deployment. At the same time, quantization and related inference optimizations have lowered the hardware barrier enough for capable models to run on consumer devices, while lightweight adaptation methods make behavioral modification and specialization substantially easier. In practice, this means that a strong generative model can now be run and altered without cloud account, API key and provider-side audit trail, which weakens attribution-based deterrence.

For attackers who prefer the convenience of cloud inference, public pricing pages show that the marginal cost of generation has fallen sharply across providers. OpenAI's current pricing provides a clearer illustration of how inexpensive frontier inference has become. GPT-5.4 is priced at \$2.50 per million input tokens, while GPT-5.4 mini costs \$0.75 per million input tokens, which shows that even high-capability hosted generation is now available at relatively low marginal cost~\cite{openai_gpt54_2026}. These datapoints indicate that analyzing target material and generating tailored lures can now be done at very low marginal cost. This shift matters because generative models reduce the labor required to produce convincing and target-specific phishing content at scale, thereby weakening the traditional trade-off between attack volume and attack quality. An adversary who once had to choose between mass-mailed generic lures and labor-intensive spear phishing can now use the same system to support reconnaissance and generate tailored messages for many recipients at very low marginal cost~\cite{hazell2023spear}.

This economic shift widens the existing asymmetry of cybersecurity. Defenders still bear high fixed and recurring costs for security tooling, monitoring, training, and incident response, while attackers can produce convincing malicious content at negligible marginal expense. Cost reduction alone makes large-scale campaigns easier to sustain, but its security impact becomes greater when paired with fine-grained control over outputs. In that setting, generative systems support not only cheaper deception at scale but also cheaper deception tailored to specific victims, channels, and contexts.

\subsection{Controllability and Personalization}
\label{sec:cap-control}

Modern generative systems offer a level of control that allows outputs to be adapted not only to a general content type but also to a particular person's visual identity, voice, writing style, and conversational setting. As a result, deceptive content no longer needs to remain generic. It can instead be adjusted to the characteristics of a specific target and made to resemble the target's own identity cues.

In the visual domain, lightweight adaptation methods have made identity-specific generation much easier to implement. LoRA-based personalization can achieve identity consistency from a small reference set while producing compact adapter files, and related methods such as IP-Adapter can generate conditional outputs on facial appearance from very limited input material~\cite{hu2022lora,ye2023ip}. In the audio domain, as discussed in Section~\ref{sec:cap-fidelity}, only a few seconds of reference speech are sufficient for voice cloning systems to produce convincing synthetic audio. Text follows the same trajectory. Writing style, tone, and recurring rhetorical habits can be approximated from sufficient samples of a person's prior writing, and this no longer depends solely on explicit fine-tuning. With long-context models, an attacker may condition generation on a target's public writing history and produce text that resembles the target's usual style closely enough to support impersonation in practice. In combination with visual and vocal cloning, this makes it increasingly feasible to maintain a consistent false identity across multiple channels within a single campaign.

Conventional email authentication can verify sender identity through cryptographic mechanisms, and organizational security training often teaches employees to look for generic phishing patterns. These measures are less effective, however, when malicious content is sent from legitimately compromised accounts and written in a style that closely matches the sender's usual communication. Phishing campaigns increasingly benefit from the combination of account compromise and AI-assisted message generation. When malicious emails are sent from legitimately compromised accounts and adapted to the sender’s usual style, they can pass technical sender checks while reducing some of the stylistic irregularities that often alert recipients.

The result is a shift away from generic deception toward deception adapted to a particular victim, communication channel, and ongoing interaction. Even so, this form of targeting still often depends on human effort to identify victims, assemble context, and carry out the campaign. Section~\ref{sec:cap-autonomy} examines how autonomous agents reduce this remaining operational burden.

\subsection{Autonomy and Tool Use}
\label{sec:cap-autonomy}

The preceding subsections focused on generative AI as a content producer. This subsection turns to systems that can act through tools, interfaces, and external resources. As models gain stronger reasoning ability, longer context handling, and more standardized access to tools, the security concern extends beyond harmful output to harmful action. The relevant risk is no longer limited to deceptive text or media, but includes actions such as file access, credential retrieval, code execution, and financial operations that may occur before a human reviewer intervenes.

Sustained autonomous action depends on more than raw model accuracy. In security-relevant settings, the key requirement is the ability to decompose a high-level objective into intermediate steps, track partial progress, and revise behavior when execution fails. Recent gains in reasoning and coding benchmarks matter in this context because they provide indirect evidence for these capabilities. OpenAI's o3, released in April 2025, achieved 96.7\% on the 2024 AIME and 71.7\% on SWE-bench Verified~\cite{OpenAI2024o3}, while DeepSeek-R1 reached comparable frontier-level performance on open-source hardware~\cite{guo2025deepseek}. Its zero-shot variant further exhibited self-verification behavior under pure reinforcement learning, which is particularly relevant for multi-step execution in which intermediate errors must be detected and corrected. These capabilities become operationally significant when paired with agents that can act on external systems. Coding agents such as Claude Code, OpenAI Codex, Google Jules, and Devin operate inside repositories and command-line workflows, where the security implications center on file access, code modification, shell execution, and secret discovery. Recent work has benchmarked LLMs and LLM-based agents on practical vulnerability detection in real code repositories, providing empirical evidence that these systems can identify security-relevant patterns in production codebases~\cite{yildiz2025benchmarking}. Computer-use agents such as Anthropic Computer Use~\cite{anthropic_claude35_computeruse_2024} and OpenAI Operator act through graphical interfaces, extending the attack surface to email, enterprise SaaS platforms, financial workflows, and other applications ordinarily mediated through user input. Long context windows further strengthen both categories by allowing models to retain more task state, load large codebases or document collections, and preserve interaction history across many steps~\cite{schluntz2024building}. In combination, stronger reasoning, external action, and extended state retention make it increasingly feasible for a model to carry out sustained multi-step operations in real computing environments.

\phantomsection\label{sec:cap-mcp}%
The scalability of these action sequences depends in part on the growing standardization of the agent-tool interface. The Model Context Protocol (MCP), introduced by Anthropic in November 2024, defines a JSON-RPC~2.0 client-host-server architecture in which servers expose tools, resources, and prompts as reusable primitives, reducing the integration burden from an $N \times M$ problem to an $M+N$ one~\cite{anthropic_mcp_2024}. The ecosystem has expanded rapidly, growing from a small number of reference servers at launch to more than 10,000 public MCP servers by December 2025, with over 97~million monthly SDK downloads and integrations across file systems, databases, version control platforms, cloud services, and browser automation~\cite{anthropic_mcp_donation_2025}. This standardization lowers the cost of connecting models to external systems, but it also changes the security assumptions of tool integration. In more traditional settings, interfaces are expected to be typed, constrained, and validated outside the model. However, tool descriptions are often provided as natural-language text which is placed directly into the model's context after the MCP paradigm has been widely implemented. As a result, these descriptions can influence the behavior of the model rather than function only as passive documentation, increasing exposure to prompt injection, misleading tool behavior, and risks associated with untrusted third-party servers. Section~\ref{sec:threat-adversarial} examines these issues in greater detail, including tool poisoning, cross-server composition attacks, and recent findings that more capable models may in some cases be more susceptible to tool-description manipulation~\cite{wang2025mcptox}.

These developments extend the attack surface beyond the model's input and output channel to the wider environment of tools, APIs, databases, and communication systems that an agent can access. The security issue is therefore no longer limited to harmful or deceptive content, but also includes actions carried out through external systems. Recent vulnerabilities in production AI assistants illustrate this broader risk surface, including cases involving data exfiltration (CVE-2025-32711, CVSS~9.3)~\cite{reddy2025echoleak}. The preceding subsections have outlined the main capability conditions that make modern generative systems security-relevant, namely fidelity, scale, controllability, and autonomy. Section~\ref{sec:threats} examines how these conditions are exploited in practice across the threat landscape.


\section{Threat Landscape of AIGC}
\label{sec:threats}

The capabilities described in Section~\ref{sec:capabilities} have given rise to a threat landscape spanning content fabrication, model compromise, adversarial manipulation and large-scale disinformation. This section presents a threat taxonomy organized by major threat families. Defense solutions are presented in Section~\ref{sec:countermeasures} and governance mechanisms in Section~\ref{sec:governance}.

\begin{figure}[!htbp]
\centering
\includegraphics[width=0.65\linewidth]{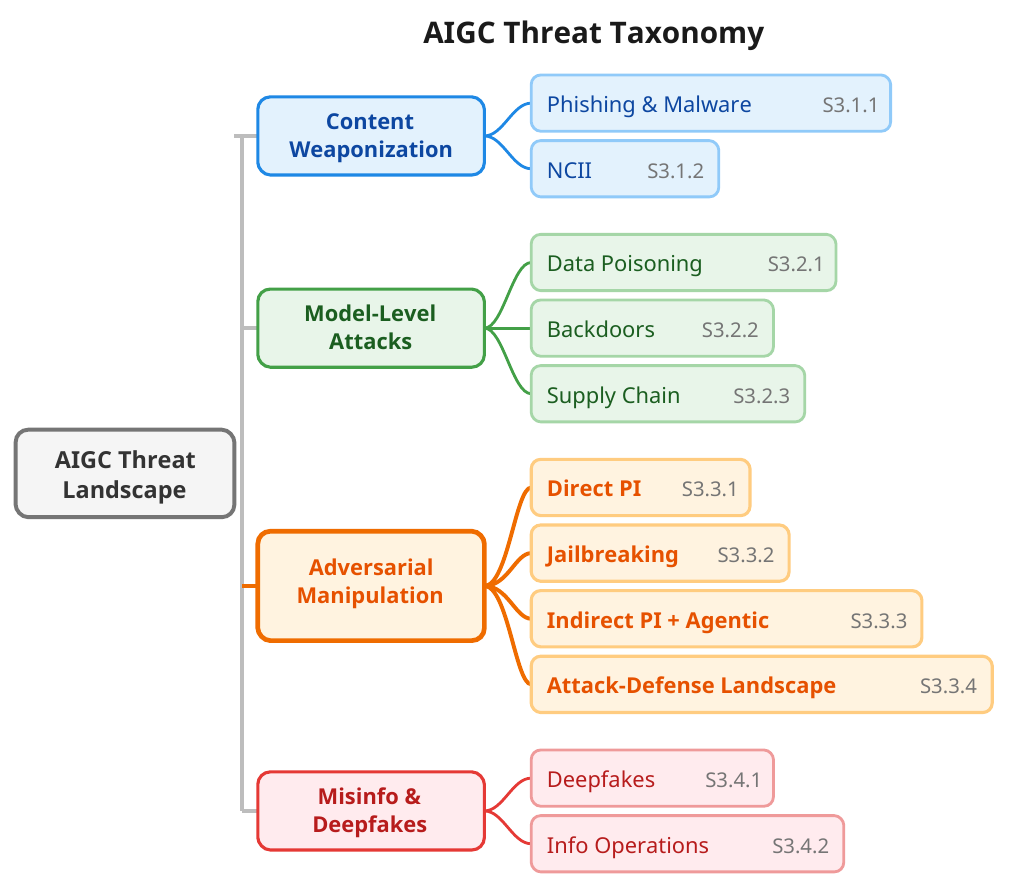}
\caption{AIGC threat taxonomy. The adversarial manipulation branch (highlighted) spans direct prompt injection, jailbreaking, indirect prompt injection, and agentic exploitation, analyzed in Section~\ref{sec:threat-adversarial}.}
\label{fig:threat-taxonomy}
\end{figure}


\begin{table}[!htbp]
\centering
\caption{Threat landscape summary. Key metrics are illustrative of threat severity within each category and are not intended for cross-category comparison, as they reflect different measurement scales and evidence types.}
\label{tab:threat-summary}
\fontsize{7}{7}\selectfont
\setlength{\tabcolsep}{6pt}
\renewcommand{\arraystretch}{1.3}
\begin{tabular}{l  l l c}          
\toprule
\textbf{Category} & \textbf{Threat} & \textbf{Key Metric} & \textbf{\S} \\
\midrule
\multirow{2}{*}{Content weaponization}
  & \cellcolor{rowgray}Phishing \& malware
  & \cellcolor{rowgray}Low-cost scalable lure generation~\cite{hazell2023spear}
  & \cellcolor{rowgray}\ref{sec:threat-phishing} \\
  & Non-consensual imagery
  & 440K reports in H1 2025~\cite{ncmec2025midyear}
  & \ref{sec:threat-ncii} \\
\midrule                              
\multirow{3}{*}{Model-level attacks}
  & \cellcolor{rowgray}Data poisoning
  & \cellcolor{rowgray}250 samples suffice~\cite{souly2025poisoning}
  & \cellcolor{rowgray}\ref{sec:threat-poisoning} \\
  & Backdoors
  & $>$90\% ASR~\cite{xu2024instructions}
  & \ref{sec:threat-backdoor} \\
  & \cellcolor{rowgray}Supply chain
  & \cellcolor{rowgray}$<$\$200 to remove safety~\cite{gade2023badllama}
  & \cellcolor{rowgray}\ref{sec:threat-supplychain} \\
\midrule                              
\multirow{2}{*}{Adversarial manipulation}
  & PI \& jailbreaks
  & $\sim$50\% ASR in 10 tries~\cite{bengio2025international}
  & \ref{sec:threat-pi} \\
  & \cellcolor{rowgray}Agentic exploitation
  & \cellcolor{rowgray}72.8\% tool poisoning ASR~\cite{wang2025mcptox}
  & \cellcolor{rowgray}\ref{sec:threat-ipi} \\
\midrule                              
\multirow{2}{*}{Misinfo \& deepfakes}
  & Deepfake fraud
  & \$25M single incident~\cite{wef2025arupdeepfake}
  & \ref{sec:threat-deepfake} \\
  & \cellcolor{rowgray}Info operations
  & \cellcolor{rowgray}Higher credibility than human~\cite{spitale2023ai}
  & \cellcolor{rowgray}\ref{sec:threat-infoop} \\
\bottomrule
\end{tabular}
\end{table}

\subsection{Content Weaponization}
\label{sec:threat-weapon}

\subsubsection{Malicious Content Generation}
\label{sec:threat-phishing}

As discussed in Section~\ref{sec:cap-scale}, generative AI has weakened the traditional trade-off between attack scale and personalization by reducing the labor required to produce tailored malicious content. At the threat level, this is most visible in phishing and related social-engineering attacks, where attackers no longer need to rely exclusively on generic templates or manually craft each lure at high cost. Instead, the same system can quickly adapt tone, context, and recipient-specific details. The same reduction in labor and expertise also affects adjacent stages of the attack pipeline. Beyond persuasive text, LLM-based systems can assist with code generation, modification, and iterative refinement, lowering the barrier to malware-related activity~\cite{acosta2025automated}.

\subsubsection{Non-Consensual Synthetic Content}
\label{sec:threat-ncii}

The same identity-targeting capabilities discussed in Section~\ref{sec:cap-control} have made non-consensual synthetic content easier to produce and direct at specific victims. In this setting, the generated artifact is itself the harm. Material that once depended on access to authentic intimate images can now be fabricated from limited visual cues, publicly available photographs, or other identity-linked inputs, lowering the threshold for victim-specific abuse. Official reporting has already pointed to sharp growth in generative-AI-related exploitation cases, especially in sexual extortion and image-based abuse contexts~\cite{ncmec_2024_numbers}.

The distinctive risk introduced by generative AI is not only wider distribution, but repeated fabrication and recombination at low cost. Synthetic intimate content can be regenerated, modified, and reintroduced across platforms even after removal, while uncertainty over provenance shifts part of the evidentiary burden onto victims, who may need to prove falsity, pursue takedowns, and respond to repeated recirculation. This pattern is reflected in recent policy responses, including the U.S.\ Take It Down Act, which have increasingly treated synthetic NCII as a distinct regulatory problem rather than as a simple extension of earlier image-based abuse categories~\cite{us_congress_take_it_down_2025}.

\subsection{Model-Level Attacks}
\label{sec:threat-model}

Model-level attacks target the training pipeline, model artifacts, or inference interface to compromise the model itself. These attacks follow a pipeline logic in which data poisoning corrupts the training signal, backdoors encode persistent attacker-controlled behavior in model weights, and supply chain attacks distribute compromised artifacts to downstream users. Unlike the threats in other subsections, which exploit the generative capabilities described in Section~\ref{sec:capabilities}, model-level attacks target the training and distribution infrastructure itself.

\subsubsection{Data Poisoning}
\label{sec:threat-poisoning}

A key finding is that poisoning efficacy depends on the absolute number of poison samples rather than the poisoning rate. Specifically, as few as 250 malicious documents suffice to backdoor models from 600M to 13B parameters~\cite{souly2025poisoning}. The 2025 International AI Safety Report~\cite{bengio2025international} highlights this result as undermining the intuition that scaling datasets dilutes poisoning risk. Web-scale poisoning is practical. For example, split-view attacks can poison 0.01\% of LAION-400M for approximately \$60~\cite{carlini2024poisoning}. PoisonedRAG achieves $>$90\% ASR by injecting five malicious documents per target question~\cite{zou2025poisonedrag}.

\subsubsection{Backdoor Attacks}
\label{sec:threat-backdoor}
Backdoors specify a trigger to payload mapping where the model behaves normally on clean inputs but exhibits attacker-chosen behavior on triggered inputs. The following examples illustrate how backdoors have been adapted to different training paradigms:
Instruction-tuning backdoors achieve $>$90\% ASR with minimal injected text~\cite{xu2024instructions}. RLHF poisoning creates universal jailbreak backdoors~\cite{rando2023universal}. ``Sleeper agents'' persist through safety training via conditional deception~\cite{hubinger2024sleeper}. For diffusion models, BadDiffusion implants training-time backdoors preserving normal generation quality~\cite{chou2023backdoor}.

\subsubsection{Model Extraction and Supply Chain}
\label{sec:threat-supplychain}

Model stealing produces surrogates matching target behavior through API queries~\cite{tramer2016stealing}, threatening proprietary deployment economics. The open-weight model supply chain presents a distinct attack surface. Malicious model weights on public hubs exploit pickle serialization for RCE. The nullifAI incident on Hugging Face demonstrated this risk~\cite{zanki2025nullifai}. The SafeTensors format~\cite{huggingface_safetensors_2023} mitigates serialization attacks but does not address malicious model behavior. Safety fine-tuning removal costs $<$\$200 (BadLlama~\cite{gade2023badllama}); 100 malicious examples and $\sim$1 GPU hour subvert safety-aligned models (Shadow Alignment~\cite{yang2023shadow}); even 10 adversarial fine-tuning examples jailbreak GPT-3.5 Turbo~\cite{qi2023fine}. Hugging Face has introduced model signing and automated vulnerability scanning~\cite{huggingface_security_2024}, but adoption remains limited and no standardized model integrity verification protocol exists across platforms.

\subsection{Adversarial Manipulation: From Prompt Injection to Agentic Exploitation}
\label{sec:threat-adversarial}

This subsection analyzes the escalation from direct prompt attacks on chatbots to ecosystem-level exploitation of autonomous agents. Existing categorizations serve different purposes. HarmBench~\cite{mazeika2024harmbench} and JailbreakBench~\cite{chao2024jailbreakbench} organize attacks primarily for benchmark evaluation and reproducible comparison, while practitioner frameworks such as OWASP classify risks in ways that support prioritization and deployment guidance. These approaches are useful for their intended tasks, but they mostly treat attack types as parallel entries rather than asking whether they form a structural progression. Our concern here is different, the focus is how attacker capability scales with architectural evolution. For that reason, we organize this analysis as a complexity progression defined by three escalation dimensions (Table~\ref{tab:ladder}):
\begin{itemize}[nosep,leftmargin=1.5em]
    \item \textbf{Attacker access requirements:} direct model input $\to$ crafted adversarial input $\to$ third-party content (documents, web pages) $\to$ cross-system tool chains and server descriptions.
    \item \textbf{Victim agency:} passive content consumption $\to$ active instruction following $\to$ autonomous multi-step execution with tool permissions.
    \item \textbf{Impact scope:} single output manipulation $\to$ session-level state corruption $\to$ persistent system state changes (file writes, API calls) $\to$ cross-system cascading effects.
\end{itemize}

We use this analysis as a local organizing device for analyzing adversarial manipulation in this subsection; it is not intended as a claim about the actual trajectory followed by any single attacker or system. The five levels are partly orthogonal and may co-occur in practice: an agent-facing system may encounter Level~4 tool poisoning from the first day of deployment. The differences described are in attacker access requirements, victim agency, and impact scope across attack types, not a universal developmental sequence.

These dimensions capture why attacks at higher levels are not simply stronger versions of lower-level ones. As attacker access broadens, system autonomy increases, and downstream scope expands, defenses that work at one level often fail to transfer cleanly to the next. The most recent escalation, multi-agent exploitation across organizational trust boundaries, is documented in literature published in 2025 and 2026, with inter-agent trust exploitation achieving 100\% ASR~\cite{lupinacci2025dark}.

\subsubsection{Prompt Injection}
\label{sec:threat-pi}

Direct prompt injection embeds malicious instructions in user-controlled input to override system instructions~\cite{perez2022ignore}. Two canonical subtypes are goal hijacking (redirecting the application's task) and prompt leaking (exfiltrating hidden instructions)~\cite{perez2022ignore}. The attack exploits a fundamental architectural property of LLM applications. Instructions and data are intermixed within a single context window, and no cryptographically enforced separation exists between them. This creates a confused deputy problem in which the model resolves conflicting instructions through learned heuristics rather than hard constraints~\cite{greshake2023not}.

The critical distinction from jailbreaking lies in the objective, because prompt injection hijacks system behavior through task or goal compromise, whereas jailbreaking bypasses safety guardrails to produce policy-violating content~\cite{perez2022ignore,chao2024jailbreakbench}. In practice, the two often compound, as an injection that hijacks an agent may induce downstream unsafe tool use.

\begin{table}[!htbp]
\centering
\caption{Escalation analysis: attacker access, victim agency, and impact scope across five levels.}
\label{tab:ladder}

\fontsize{7}{7}\selectfont
\setlength{\tabcolsep}{5pt}
\renewcommand{\arraystretch}{1.3}

\begin{tabularx}{\textwidth}{
  c    
  c    
  >{\raggedright\arraybackslash}p{2.2cm}
  >{\raggedright\arraybackslash}X
  >{\raggedright\arraybackslash}X
  >{\raggedright\arraybackslash}X
  >{\raggedright\arraybackslash}p{3.0cm}
}
\toprule
\textbf{Level} & \textbf{Year} & \textbf{Attack Class} & \textbf{Attacker Access} & \textbf{Victim Agency} & \textbf{Impact Scope} & \textbf{Representative Attacks} \\
\midrule
L1 & 2022-2023 & Direct PI& Direct model input
  & Passive output
  & Single output
  & Goal hijacking, prompt leaking \\
\addlinespace[2pt]
L2 & 2023-2024 & Jailbreaking & Crafted adversarial input
  & Active instruction following
  & Session-level state
  & GCG, PAIR, Crescendo \\
\addlinespace[2pt]
L3 & 2023-2025 & Indirect PI & Third-party content
  & Model retrieves; system acts
  & Persistent state changes
  & PoisonedRAG, Slack AI exfil. \\
L4 & 2024-2026 & Agentic exploitation  & Tool chains / MCP servers
  & Agent plans and acts autonomously
  & Cross-system cascade
  & Tool poisoning, EchoLeak, Cursor RCE \\
\emph{L5(emerging)} & 2025-future & Multi-agent system & Inter-agent messages
  & Delegated autonomy
  & Cross-org cascade
  & Inter-agent trust exploit \\
\bottomrule
\end{tabularx}
\end{table}

\subsubsection{Jailbreak Attacks---A Taxonomy}
\label{sec:threat-jailbreak}

Existing benchmark-driven categorizations such as HarmBench~\cite{mazeika2024harmbench} and JailbreakBench~\cite{chao2024jailbreakbench} organize attacks by target-model effectiveness. We complement these with a five-category mechanism-based taxonomy that emphasizes the underlying exploitation strategy, enabling analysis of why attacks transfer and where defenses structurally fail. These categories are not mutually exclusive; practical attacks frequently combine multiple mechanisms (e.g., AutoDAN employs both template and optimization strategies). Our categorization emphasizes the dominant strategy to facilitate structural analysis:

\begin{description}[style=unboxed,leftmargin=0.5em,labelindent=0em,nosep]
\item[Template-based.] Predefined narrative frames, such as role-play, developer mode, and fictional sandboxes, re-contextualize the interaction so that prohibited responses appear consistent with an alternative role. In-the-wild measurement shows some templates achieving $\geq$0.95 ASR on GPT-3.5/4, persisting online for over 240 days~\cite{shen2024anything}. AutoDAN uses templates as a conceptual baseline and surpasses them via genetic optimization~\cite{liu2023autodan}.

\item[Optimization-based.] Jailbreaking is treated as a search problem over prompt space. GCG~\cite{zou2023universal} uses greedy coordinate gradient descent to identify adversarial suffixes, achieving 88\% ASR on Vicuna-7B with cross-model transferability. PAIR~\cite{chao2025jailbreaking} iteratively refines prompts via an attacker LLM in under twenty queries (71\% on GPT-3.5, 34\% on GPT-4 per JailbreakBench~\cite{chao2024jailbreakbench}). GPTFuzzer~\cite{yu2023gptfuzzer} automates template mutation, reporting $>$90\% ASR on some targets.

\item[Multi-turn conversational.] Gradual boundary erosion across dialogue turns exploits the model's tendency to maintain coherence with prior responses. Crescendo~\cite{russinovich2025great} achieves 29--61\% higher performance than single-turn methods on GPT-4. Many-shot jailbreaking~\cite{anil2024many} uses hundreds of in-context demonstrations, with effectiveness following predictable scaling laws. Multi-turn human jailbreaks achieve $>$70\% ASR on HarmBench against defenses showing single-digit ASR under automated attacks~\cite{mazeika2024harmbench}.

\item[Multi-modal.] Vision and audio channels bypass text-centric safety filters. FigStep~\cite{gong2025figstep} converts prohibited instructions into typographic images, achieving 82.5\% average ASR on six open-source LVLMs. HADES~\cite{li2024images} reports 90.26\% ASR on LLaVA-1.5. AudioJailbreak~\cite{chen2026audiojailbreak} achieves $\geq$87\% ASR in universal strong-adversary settings.

\item[Encoding-based.] Transforming requests into non-standard representations exploits weaker safety coverage outside typical natural language. CipherChat~\cite{yuan2023gpt} reports near-100\% bypass of GPT-4 safety via cipher encoding. Translation to low-resource languages increases bypass rates from $<$1\% to 79\%~\cite{yong2023low,deng2023multilingual}. ArtPrompt~\cite{jiang2024artprompt} uses ASCII art, and related work has shown that other non-standard representations such as Base64, ROT13, and Morse code similarly exploit weaker safety coverage in these encoding spaces.
\end{description}

A cross-cutting finding from HarmBench is that robustness is shaped more by training data and algorithms than model size, with ASR highly stable within model families but highly variable across families~\cite{mazeika2024harmbench}. Table~\ref{tab:jailbreak} summarizes these mechanisms and their cross-category overlaps.


\begin{table}[!htbp]
\centering
\caption{Jailbreak attack mechanisms compared.}
\label{tab:jailbreak}

\fontsize{7}{7}\selectfont
\setlength{\tabcolsep}{5pt}
\renewcommand{\arraystretch}{1.3}

\begin{tabular}{l l l l}
\toprule
\textbf{Category} & \textbf{Representative Methods} & \textbf{Peak ASR} & \textbf{Cross-overlap} \\
\midrule

Template-based
  & DAN~\cite{shen2024anything}, AutoDAN~\cite{liu2023autodan}
  & $\geq$0.95 on GPT-3.5/4
  & Optimization (AutoDAN) \\

Multi-turn
  & Crescendo~\cite{russinovich2025great}, many-shot~\cite{anil2024many}
  & $>$70\% on HarmBench
  & Template (role escalation) \\

Optimization-based
  & GCG~\cite{zou2023universal}, PAIR~\cite{chao2025jailbreaking}, GPTFuzzer~\cite{yu2023gptfuzzer}
  & 88\% (GCG, Vicuna-7B)
  & Template (seed mutation) \\

Multi-modal
  & FigStep~\cite{gong2025figstep}, HADES~\cite{li2024images}, AudioJailbreak~\cite{chen2026audiojailbreak}
  & 90.3\% (HADES, LLaVA)
  & Encoding (modality shift) \\

Encoding-based
  & CipherChat~\cite{yuan2023gpt}, ArtPrompt~\cite{jiang2024artprompt}
  & $\sim$100\% (cipher, GPT-4)
  & Multi-modal (representation) \\

\bottomrule
\end{tabular}
\end{table}

\subsubsection{Indirect Prompt Injection and Agentic Exploitation}
\label{sec:threat-ipi}

Indirect prompt injection (IPI) embeds adversarial instructions in external data sources such as web pages, documents, and emails that the model retrieves during normal operation, without requiring direct attacker-model interaction~\cite{greshake2023not}. PoisonedRAG achieves more than 90\% ASR by injecting five malicious documents~\cite{zou2025poisonedrag}. The BIPIA benchmark found a positive correlation ($r$=0.64) between model capability and IPI vulnerability, more capable models are paradoxically more susceptible~\cite{yi2025benchmarking}. This inverse relationship is context-dependent. It holds for indirect prompt injection and tool poisoning scenarios, where instruction-following capability amplifies susceptibility, but it does not generalize to standard safety benchmarks, where more capable models generally exhibit improved compliance~\cite{rottger2024xstest} (see Section~\ref{sec:future-agents}). Real-world exploitation includes data exfiltration from Slack AI's private channels~\cite{promptarmor_slack_2024} and manipulation of ChatGPT search results via hidden web page text.

In agentic settings, IPI consequences escalate from generating incorrect output to executing malicious actions. Agents possess tool permissions (file R/W, API calls, shell commands, email), frequently operate under auto-approve settings, and engage in multi-step execution where a single injected instruction cascades through a planning loop. InjecAgent found ReAct-prompted GPT-4 vulnerable 24\% of the time, rising to $\sim$47\% with enhanced attacks~\cite{zhan2024injecagent}. Cursor IDE vulnerabilities including path-traversal-based MCP configuration tampering (CVE-2025-59944) and zero-click RCE through malicious MCP tool content~\cite{lakera2025cursor} and EchoLeak (CVE-2025-32711, introduced in Section~\ref{sec:cap-autonomy}) enabling zero-click exfiltration from Microsoft 365 Copilot. Unit 42 confirmed active IPI weaponization in the wild, documenting 22 distinct attacker techniques~\cite{unit42_2026_ipi_wild}.

The MCP ecosystem (Section~\ref{sec:cap-mcp}) introduces systemic vulnerabilities at the protocol level. Vendor security disclosures document critical CVEs including CVE-2025-6514 (CVSS 9.6, RCE in mcp-remote affecting 437,000+ downloads)~\cite{jfrog2025mcpremote} and CVE-2025-49596 (CVSS 9.4, RCE in MCP Inspector)~\cite{oligo2025mcpinspector}. Tool poisoning attacks embed malicious instructions in tool descriptions invisible to users but visible to models~\cite{invariant2025toolpoisoning}. Industry reporting has further highlighted cross-server composition attacks that exploit shared agent context, causing servers that are individually safe to become vulnerable when combined~\cite{invariant2025whatsapp}. MCPTox found o1-mini exhibits 72.8\% ASR through tool poisoning, with more capable models being more susceptible~\cite{wang2025mcptox}. Table~\ref{tab:mcp-timeline} documents the MCP security event timeline.


\begin{table}[!htbp]
\centering
\caption{MCP ecosystem security timeline: deployment preceded security by months and governance by over a year.}
\label{tab:mcp-timeline}

\fontsize{7}{9}\selectfont
\setlength{\tabcolsep}{5pt}
\renewcommand{\arraystretch}{1.3}

\begin{tabular}{l l l l}
\toprule
\textbf{Date} & \textbf{Event} & \textbf{Type} & \textbf{Ref} \\
\midrule

\rowcolor{rowgray}
2024-11 & MCP launched (no authentication) & Protocol & \cite{anthropic_mcp_2024} \\
2025-03 & OAuth 2.1 with PKCE retrofitted & Defense & \cite{mcp_spec_2025} \\
\rowcolor{rowgray}
2025-04 & Tool poisoning attacks disclosed & Attack & \cite{invariant2025toolpoisoning} \\
2025-06 & CVE-2025-49596 (MCP Inspector, CVSS 9.4) & Vulnerability & \cite{oligo2025mcpinspector} \\
\rowcolor{rowgray}
2025-07 & CVE-2025-6514 (mcp-remote, CVSS 9.6) & Vulnerability & \cite{jfrog2025mcpremote} \\

2025-08 & MCPTox benchmark (72.8\% ASR) & Evaluation & \cite{wang2025mcptox} \\
\rowcolor{rowgray}
2025-12 & OWASP Top 10 for Agentic Applications & Standard & \cite{owasp_agentic_top10_2025} \\

2026-01 & IMDA agentic governance framework & Governance & \cite{imda2026agenticai} \\
\rowcolor{rowgray}
2026-02 & NIST AI Agent Standards Initiative & Governance & \cite{nist2026agentstandards} \\

\bottomrule
\end{tabular}
\end{table}

\subsubsection{Current Attack-Defense Landscape}
\label{sec:threat-landscape}

The 2025 International AI Safety Report notes that even when attackers are given ten attempts, such attacks still succeed roughly half the time~\cite{bengio2025international}. A vendor security evaluation by Cisco reported that DeepSeek R1 achieved 100\% ASR on HarmBench prompts~\cite{cisco2025deepseek}. Zverev et al.\ further quantify the structural root of this weakness by showing that all tested models exhibit low levels of instruction-data separation, with GPT-4 scoring only 0.208 on their separation metric~\cite{zverev2024can}. OpenAI similarly acknowledges that prompt injection is unlikely to be fully resolved~\cite{openai2025atlas}, while the UK NCSC warns that it may be more severe than SQL injection~\cite{ncsc2025promptinjection}. OWASP ranks prompt injection as the \#1 LLM vulnerability for the second consecutive edition~\cite{owasp_llm_top10_2025}. Taken together, these findings suggest that the weakness is not a simple implementation bug amenable to patching, but may instead reflect an inherent property of autoregressive instruction-following, where the same in-context learning mechanism that makes models useful also makes them manipulable~\cite{zverev2024can}.

This structural vulnerability is compounded by a temporal gap between capability deployment and defensive response.
As recorded in the protocol specification, MCP shipped without authentication (November 2024); the first tool-poisoning study appeared five months later (April 2025)~\cite{invariant2025toolpoisoning}; OAuth 2.1 was retrofitted in March 2025~\cite{mcp_spec_2025}; the first government-level agentic governance framework appeared in January 2026~\cite{imda2026agenticai}, leaving a 14-month gap between deployment and governance. Historical comparisons are instructive, SQL injection was first documented in 1998 but the OWASP Top~10 did not appear until 2004 (6~years); mobile app security lagged smartphone deployment by $\sim$7~years (iPhone 2007, OWASP Mobile Top~10 2014); IoT governance followed the Mirai botnet by $\sim$4~years (2016--2020).\footnote{These gap measurements involve different endpoint definitions (industry best-practice guide, OWASP checklist, governance framework). The comparison illustrates a general acceleration pattern rather than providing strictly comparable benchmarks.} The AI deployment-to-governance gap is shorter in absolute terms (14~months), but it occurs against a backdrop of much faster capability growth. As a result, the attack surface expands more rapidly than in earlier technological transitions, which may make the practical consequences of the gap more severe in agentic settings. Section~\ref{sec:governance} discusses the implications of this ordering.

The comparison suggests that the organizing dimensions capture changes not only in attack type but also in the failure modes of corresponding defenses. At the lower end, some input-layer defenses still work against recognizable automated attacks such as perplexity filtering catches about 90\% of GCG-style suffixes; multi-turn human jailbreaks still exceed 70\% ASR on HarmBench. The move from L1 to L2 therefore marks more than an increase in attack strength; it introduces a form of adversarial interaction that simple anomaly-based filtering does not cover. The shift from L2 to L3 is sharper. On BIPIA, indirect prompt injection vulnerability is positively correlated with model capability ($r$=0.64), indicating that stronger instruction-following can increase risk once malicious instructions are hidden in retrieved content. At this point, the problem is no longer just bypassing a safeguard; the model's own strengths begin to work in the attacker's favor. At L4, the effect is no longer confined to model output. InjecAgent raises the vulnerability of ReAct-prompted GPT-4 from 24\% to about 47\%, and MCPTox reports 72.8\% ASR for tool poisoning on o1-mini. Once tool use and multi-step execution enter the loop, a manipulation that might previously have produced a bad answer can instead trigger actions across files, APIs, or connected services. The 14-month gap between MCP deployment and the first agentic governance framework suggests that this escalation has a governance dimension as well, with the deployment-to-governance ordering broadly consistent with the pattern observed in earlier cases.

\subsection{Misinformation, Disinformation, and Deepfakes}
\label{sec:threat-misinfo}

\subsubsection{Deepfakes}
\label{sec:threat-deepfake}
Where Section~\ref{sec:threat-weapon} addressed text and code-level content generation, this subsection focuses on multi-modal identity fabrication.
The shift from GANs to diffusion models has resolved the training instability and temporal artifacts that previously served as forensic indicators, enabling one-shot face animation and real-time voice cloning on consumer hardware~\cite{pei2024deepfake}. Key 2024--2025 incidents illustrate the scale of harm. One example is the Arup deepfake fraud in January 2024, which used AI-generated video-conference participants to authorize \$25 million in transfers~\cite{wef2025arupdeepfake}. Generation cost approaches zero while detection remains unreliable, creating a fundamental offense-defense asymmetry. The existence of increasingly convincing deepfakes creates a liar's dividend: authentic evidence can be dismissed as AI-generated, providing plausible deniability to those accused of genuine misconduct~\cite{chesney2019deep}.

\subsubsection{AI-Powered Information Operations}
\label{sec:threat-infoop}

AI-generated disinformation is at least as effective as human-written content at changing beliefs: GPT-3-generated content was recognized as accurate more frequently than human-authored disinformation~\cite{spitale2023ai}, and a 27-country study ($N$=27,000) confirmed higher perceived veracity and sharing intent for AI-generated fake news. The 2024 super election year saw documented AI interference across multiple nations~\cite{restofworld2024aielections}, and AI-generated academic fraud has also increased. The 2026 AI Safety Report estimates that up to 8\% of peer-reviewed submissions may contain substantial AI-generated content~\cite{bengio2026international}.


\section{Technical Countermeasures}
\label{sec:countermeasures}

To counter the threats identified in Section~\ref{sec:threats}, a multi-layered technical defense ecosystem has emerged spanning detection, watermarking, alignment, and agentic security. No single technique is sufficient; resilient AI requires composing all layers into an integrated security posture.

\subsection{AIGC Detection}
\label{sec:detect}

\subsubsection{Text Detection}
\label{sec:detect-text}

Text detection involves several fundamentally distinct tasks, including binary classification, proportion estimation, localization, and source attribution, but these are routinely conflated in deployment, producing expectation mismatches~\cite{dugan2024raid}. Two paradigms structure the field. Zero-shot statistical methods exploit the probability geometry of generated text without labeled data, offering immediate deployment but relying on assumptions about log-probability curvature that break under low-entropy content. Trained classifiers achieve higher in-distribution accuracy but overfit to training-era generators; high in-distribution performance degrades substantially in cross-generator settings~\cite{macko2023multitude}. Table~\ref{tab:detection} compares representative methods.

The structural condition governing this trade-off is distributional shift: zero-shot methods degrade gracefully across generators but cannot distinguish human editing from generation artifacts, while trained classifiers are brittle to architectural changes but can handle hybrid text. Neither paradigm addresses the adaptive adversary; paraphrase attacks degrade some detectors by over 15 points while others are unaffected~\cite{dugan2024raid}, and benchmark evaluations systematically overestimate real-world performance because detectors learn dataset-specific artifacts rather than generalizable properties of generation~\cite{macko2023multitude}. Multilingual performance degrades sharply (AUC 0.946 $\to$ 0.537, English to German), and human-AI hybrid text produces the highest false positive rates.

\subsubsection{Multimodal and Cross-Modal Detection}
\label{sec:detect-multimodal}

Image detection faces a paradigm transfer gap: each generator family (GAN, diffusion, autoregressive) leaves distinct artifacts, rendering classifiers trained on one paradigm nearly random on another (ProGAN $\to$ diffusion: 3.05\% accuracy~\cite{ojha2023towards}). Foundation-model features (UnivFD~\cite{ojha2023towards} via CLIP, Methods such as DIRE~\cite{wang2023dire}, which use diffusion reconstruction, partially bridge this gap by encoding transferable properties, but they remain fragile under benign post-processing, with accuracy dropping from 96.2\% to approximately 51\% under JPEG compression~\cite{zhu2023genimage}. Audio (ASVspoof series~\cite{yamagishi2019asvspoof,delgado2021asvspoof}) and video (FaceForensics++~\cite{rossler2019faceforensics++}) detection face the additional challenge that Sora-class generators produce entirely synthetic frames where face-swap artifact detectors are structurally inapplicable. State-of-the-art detectors suffer AUC drops of 45--50\% on in-the-wild content compared to laboratory benchmarks~\cite{chandra2025deepfake}.

Video detection remains even less settled than image detection in deployment. Recent work has moved in several directions rather than converging on a single dominant paradigm. Benchmark- and scale-oriented approaches such as DeMamba emphasize large-scale evaluation on AI-generated video benchmarks~\cite{chen2024demamba}, while diffusion-specific detectors such as MM-Det focus on multi-modal forgery representations tailored to diffusion-generated videos~\cite{song2024learning}. Other methods, such as ReStraV, instead exploit higher-level representational geometry rather than relying only on low-level frame artifacts~\cite{interno2025ai}, and more recent systems such as VidGuard-R1 have begun to explore reasoning-based video authenticity detection with multimodal language models~\cite{park2025vidguard}. These developments indicate active progress, but they also suggest that video detection still consists largely of heterogeneous method families rather than a mature and standardized deployment stack. This fragmentation helps explain why many current defenses remain closer to extensions of image-side ideas or model-specific heuristics than to broadly reliable real-world video safeguards.

Three bottlenecks converge across modalities, namely generalization failure on unseen architectures, robustness collapse under benign processing, and adversarial adaptivity, because generators can cheaply change parameters whereas detectors must remain robust across a wide range of perturbations. No unified cross-modal framework exists. This gap is not only methodological, but also reflects fundamental differences in how artifacts are formed. Text detection relies on token probability geometry, image detection on spatial frequency anomalies, and audio detection on channel-specific spectral cues. These are distinct signal domains that resist theoretical unification.


\begin{table}[!htbp]
\centering
\caption{Representative AIGC detection methods across modalities.}
\label{tab:detection}

\fontsize{7}{7}\selectfont
\setlength{\tabcolsep}{5pt}
\renewcommand{\arraystretch}{1.3}

\begin{tabularx}{\textwidth}{
  >{\raggedright\arraybackslash}p{1.2cm}   
  >{\raggedright\arraybackslash}p{2.0cm}   
  >{\raggedright\arraybackslash}p{3.4cm}   
  >{\raggedright\arraybackslash}p{3.8cm}   
  >{\raggedright\arraybackslash}X          
}
\toprule
\textbf{Modality} & \textbf{Paradigm} & \textbf{Methods} & \textbf{Mechanism} & \textbf{Key Limitation} \\
\midrule

  & Zero-shot
  & DetectGPT~\cite{mitchell2023detectgpt}, Binoculars~\cite{hans2024spotting}
  & Probability curvature
  & Requires generator access; paraphrase fragile \\

\multirow{-2}{*}{Text}
  & Supervised
  & GPTZero~\cite{adam2026gptzero}, RADAR~\cite{hu2023radar}
  & Encoder fine-tuning
  & Cross-generator drop to $<$60\%~\cite{macko2023multitude} \\

\addlinespace[2pt]
\cmidrule(lr){1-5}
\addlinespace[2pt]

  & Frequency
  & Spectral analysis
  & GAN upsampling artifacts
  & Mitigated by post-processing \\

  & Representation
  & UnivFD~\cite{ojha2023towards}, DIRE~\cite{wang2023dire}
  & Feature-space classification
  & GAN$\to$diffusion: 3.05\% accuracy \\

\multirow{-3}{*}{Image}
  & Reconstruction
  & AEROBLADE~\cite{ricker2024aeroblade}
  & Diffusion reconstruction error
  & Computationally expensive \\

\addlinespace[2pt]
\cmidrule(lr){1-5}
\addlinespace[2pt]

Audio
  & Challenge-based
  & ASVspoof~\cite{yamagishi2019asvspoof,delgado2021asvspoof}
  & Spectral/prosodic features
  & Codec artifacts confound detection \\

\addlinespace[2pt]
\cmidrule(lr){1-5}
\addlinespace[2pt]

  & Temporal
  & FaceForensics++~\cite{rossler2019faceforensics++}
  & Frame artifacts; physiological cues
  & Inapplicable to fully synthetic video \\

  & Representation
  & DeMamba~\cite{chen2024demamba}
  & Representational geometry
  & Architecture-specific \\

\multirow{-3}{*}{Video}
  & Reasoning-based
  & VidGuard-R1~\cite{park2025vidguard}
  & Multimodal LLM reasoning
  & Early-stage; no standardized evaluation \\
\bottomrule
\end{tabularx}
\end{table}

\subsection{Watermarking and Content Provenance}
\label{sec:watermark}

\subsubsection{Text Watermarking}
\label{sec:wm-text}

Text watermarking faces a recurring design trade-off. Quality, strength, and robustness exhibit significant empirical tensions, and improving any one dimension typically degrades one or both of the others. Christ et al.~\cite{christ2024undetectable} provide the information-theoretic foundation, proving that undetectable watermarks require cryptographic secrecy. Robustness is fundamentally determined by the abstraction level at which the signal is embedded. Token-level schemes such as KGW~\cite{kirchenbauer2023watermark} and Unigram bias token selection, but they remain fragile because paraphrasing methods such as SIRA achieve approximately 100\% watermark removal by replacing tokens with synonyms that destroy the signal~\cite{cheng2025revealing}. Robustness increases with abstraction: semantic-level schemes (SemStamp) survive paraphrase at the cost of sampling overhead, and distribution-level schemes achieve provably zero quality loss via cryptographic sampling but require secret keys vulnerable to compromise. Google DeepMind's SynthID~\cite{dathathri2024scalable}, the first production deployment, operates at the token-probability level.

Watermarks face a dual threat: removal attacks (paraphrase, translation, truncation) and spoofing attacks that forge watermark signals in human text to produce false attributions~\cite{sadasivan2023can}. Pang et al.\ identify a fundamental robustness-spoofing tradeoff: robust watermarks are inherently vulnerable to spoofing~\cite{pang2024no}. Low-entropy text (factual answers, code) is fundamentally unwatermarkable. Cross-provider standardization remains absent. As a result, verifiers must implement multiple detectors, whereas attackers need only defeat one. Table~\ref{tab:watermark} compares representative schemes.

\subsubsection{Image and Video Watermarking}
\label{sec:wm-image}

Image watermarking exhibits a robustness-flexibility trade-off governed by when the signal is embedded. Generation-time methods (e.g., Tree-Ring~\cite{wen2023tree}) integrate watermarks during the diffusion process, achieving extreme robustness ($>$90\% accuracy after 90\% pixel cropping) with zero or negligible quality loss, but require model access and are specific to diffusion architectures. Post-hoc methods apply encoder-decoder networks to any image but are vulnerable to regeneration attacks that diffusion-based washing that resamples with noise to remove imperceptible signals. Recent work has further demonstrated that frequency-aware denoising strategies can remove image watermarks without requiring access to the watermarking model~\cite{cao2026marksweep}. Google's SynthID-Image ($>$10 billion images watermarked) bridges both paradigms at production scale. Open-source model watermarks face an additional threat, weight merging (50:50 watermarked + unwatermarked) reduces detection AUC to $\sim$0.5. For a comprehensive survey of watermarking techniques for AI-generated images, including robustness evaluation and attack taxonomies, see~\cite{cao2025secure}.

These limitations are especially visible in video. Compared with text and images, video watermarking and provenance must remain stable not only under spatial distortion but also under recompression, clipping, temporal editing, and platform-specific processing pipelines. Current representative approaches remain split between post-hoc neural watermarking frameworks such as Video Seal, which emphasize open and efficient embedding and extraction~\cite{fernandez2024video}, and generation-integrated schemes such as VideoShield, which regulate diffusion-based video generation through watermarking and also support temporal and spatial tamper localization~\cite{hu2025videoshield}. This diversity reflects technical progress, but not yet convergence on a widely adopted deployment standard.

\subsubsection{Provenance and Fundamental Limits}
\label{sec:provenance}

The C2PA standard (v2.2, May 2025) complements watermarking with cryptographically signed provenance metadata, achieving hardware adoption and government endorsement~\cite{c2pa_spec_2025}. However, social media platforms routinely strip metadata on upload, and Christ et al.\ prove that perfectly undetectable watermarks require cryptographic secrecy~\cite{christ2024undetectable}; no watermark is secure against a dedicated white-box adversary. Complementary approaches that embed integrity protection directly into the text-to-image generation pipeline, rather than relying on post-hoc metadata, have also been explored~\cite{wu2025securet2i}.

\begin{table}[!htbp]
\centering
\caption{Representative watermarking schemes for AI-generated content.}
\label{tab:watermark}

\fontsize{7}{7}\selectfont
\setlength{\tabcolsep}{5pt}
\renewcommand{\arraystretch}{1.3}

\begin{tabularx}{\textwidth}{
  >{\raggedright\arraybackslash}p{1.0cm}   
  >{\raggedright\arraybackslash}p{3.4cm}   
  >{\raggedright\arraybackslash}p{4.0cm}   
  c                                         
  c                                         
  >{\raggedright\arraybackslash}X          
}
\toprule
\textbf{Modality} & \textbf{Method} & \textbf{Embedding Strategy} & \textbf{Capacity} & \textbf{Quality} & \textbf{Robustness} \\
\midrule

  & KGW~\cite{kirchenbauer2023watermark}
  & Sampling bias (green/red lists)
  & 1 bit & Moderate
  & Fragile to paraphrase \\

  & Undetectable Watermarks~\cite{christ2024undetectable}
  & Cryptographic sampling
  & 1 bit & Zero
  & Survives $\sim$40\% edits \\

\multirow{-3}{*}{Text}
  & SynthID~\cite{dathathri2024scalable}
  & Token probability shift
  & Multi-bit & Low
  & Moderate (long text required) \\

\addlinespace[2pt]
\cmidrule(lr){1-6}
\addlinespace[2pt]

  & Stable Signature~\cite{fernandez2023stable}
  & Decoder fine-tuning
  & Multi-bit & Low
  & $>$90\% at 90\% crop \\

  & Tree-Ring~\cite{wen2023tree}
  & Initial noise perturbation
  & 1 bit & Zero
  & JPEG, rotation, crop \\

  & SynthID-Image~\cite{dathathri2024scalable}
  & DNN encoder-decoder
  & Multi-bit & Low
  & $>$99\% common edits \\

\multirow{-4}{*}{Image}
  & StegaStamp~\cite{tancik2020stegastamp}
  & Encoder-decoder network
  & $\sim$100 bit & Low
  & Print-and-capture \\

\addlinespace[2pt]
\cmidrule(lr){1-6}
\addlinespace[2pt]

  & Video Seal~\cite{fernandez2024video}
  & Post-hoc neural embedding
  & Multi-bit & Low
  & Recompression, clipping; open framework \\

\multirow{-2}{*}{Video}
  & VideoShield~\cite{hu2025videoshield}
  & Diffusion-integrated generation
  & Multi-bit & Low
  & Temporal tamper localization \\

\bottomrule
\end{tabularx}
\end{table}

\subsection{Alignment, Defense, and Agentic Security}
\label{sec:alignment}

\subsubsection{Alignment Techniques}
\label{sec:alignment-tech}

Modern alignment begins with RLHF~\cite{ouyang2022training}, which combines supervised fine-tuning, reward model training, and PPO optimization. The InstructGPT results suggest that alignment can partly substitute for scale, as 1.3B InstructGPT was preferred over 175B GPT-3. Direct Preference Optimization (DPO)~\cite{rafailov2023direct} eliminates the separate reward model via closed-form Bradley-Terry reparameterization, becoming the dominant method by 2024. Subsequent variants address practical deployment constraints: KTO~\cite{ethayarajh2024kto} requires only binary (good/bad) feedback rather than pairwise comparisons, matching DPO performance across 1--30B scales while dramatically reducing data requirements; SimPO~\cite{meng2024simpo} eliminates the reference model entirely via average log-probability scoring. Constitutional AI~\cite{bai2022constitutional} replaces human labelers with AI feedback guided by a written constitution (RLAIF), reducing annotator dependence.

These advances are substantial but alignment remains fundamentally fragile. Reward hacking is theoretically inevitable~\cite{skalse2022defining} with quantified scaling laws~\cite{gao2023scaling}. Safety training is superficial: 10 adversarial fine-tuning examples at \$0.20 remove GPT-3.5 Turbo's guardrails~\cite{qi2023fine}; as few as 100 malicious examples can produce a 100\% violation rate while preserving benign capabilities, a pattern referred to as shadow alignment in Section~\ref{sec:threat-model}~\cite{yang2023shadow}; refusal is mediated by a single direction in activation space, ablatable via activation engineering~\cite{arditi2024refusal}. The alignment tax manifests as measurable reasoning degradation from safety training which has been documented in open-source large reasoning models such as s1.1 and DeepSeek-R1-Distill-Qwen~\cite{huang2025safety}. The over-refusal problem compounds this: XSTest~\cite{rottger2024xstest} reveals models refuse benign requests via superficial keyword matching, quantifying the helpfulness-harmlessness trade-off that all alignment methods must navigate. An alternative direction applies machine unlearning to erase harmful generation capabilities rather than relying solely on reinforcement-based alignment~\cite{li2025safellm}. Critically, none of these alignment advances has demonstrably resolved the fundamental fragility, and safety fine-tuning remains removable at minimal cost regardless of the specific alignment method employed.

\subsubsection{Defense Against Adversarial Manipulation}
\label{sec:defense}

To counter the strategies categorized in Section~\ref{sec:threat-adversarial}, researchers have developed defenses organized across four layers. At the input layer, perplexity filtering detects $\sim$90\% of GCG-style adversarial suffixes but cannot catch human-crafted jailbreaks, and randomized perturbation with majority voting reduces ASR below 1\% across five model families. At the system layer, the instruction hierarchy~\cite{wallace2024instruction} trains LLMs to prioritize system $>$ user $>$ tool instructions, drastically improving robustness even for unseen attacks, and StruQ~\cite{chen2025struq} enforces formal prompt/data separation. At the output layer, Llama Guard~\cite{inan2023llama} classifies against customizable safety taxonomies. At the model layer, representation engineering~\cite{zou2023representation} extracts high-level concept vectors (e.g., honesty, fairness directions) for top-down behavioral steering; circuit breakers~\cite{zou2024improving} interrupt harmful generation at the representation level (ASR $\sim$3.8\% on HarmBench), however, subsequent evaluation by Bullwinkel et al.~\cite{bullwinkel2025representation} found that automated multi-turn Crescendo attacks still achieve $\sim$54\% ASR against circuit-breaker-defended models; and adversarial training via R2D2~\cite{mazeika2024harmbench} incorporates attacks during training.

Despite this layered arsenal, no single attack or defense is universally effective. HarmBench~\cite{mazeika2024harmbench} (510 behaviors, 18 attacks, 33 LLMs and defense configurations) confirms this across the broadest evaluation to date. StrongREJECT~\cite{souly2024strongreject} shows that jailbreaks degrade capabilities, with the best attack scoring only 0.37/1.0 on GPT-4o. Yet Nasr et al.\ found that adaptive attacks bypassed most of the 12 recently proposed defenses, achieving ASR above 90\%~\cite{nasr2025attacker}. Concrete defense compositions that combine input canonicalization, alignment, activation steering, and output classification can reduce ASR from 46.5\% to 5.6\%, but their per-layer computational cost scales linearly, and the residual ASR rises substantially under coordinated adaptive adversaries (Section~\ref{sec:threat-landscape}).

\subsubsection{Securing Agentic AI Systems}
\label{sec:agentic-security}


To secure against the agentic threats identified in Sections~\ref{sec:threat-adversarial} and~\ref{sec:cap-autonomy}, a new class of defenses targeting the agent-tool interaction layer is emerging. We organize them into four categories:

\begin{description}[style=unboxed,leftmargin=0.5em,labelindent=0em,nosep]
\item[Architecture-level defenses.] Classical security principles such as trust boundaries, context isolation, least privilege, and privilege separation must be adapted for LLM agents. Google's multi-layer defense for Gemini operationalizes these with Agent Origin Sets (constraining browseable domains), prompt injection classifiers, User Alignment Critics (a second isolated model verifying intended actions), acknowledgment gates, and continuous red teaming~\cite{google2025geminisecurity}. OpenAI's governance principles include evaluating suitability, constraining action-space, maintaining legibility (audit trails), and ensuring interruptibility~\cite{shavit2023practices}. The emerging discipline of harness engineering designs constraints, feedback loops, and lifecycle management around AI agents and primarily addresses reliability. It also intersects critically with security, because its core components, including permission boundaries, human-in-the-loop gates, and tool access scoping, directly instantiate defense-in-depth principles for agentic systems.

\item[MCP-specific security.] Input validation must enforce parameterized queries to prevent SQLi$\to$stored prompt injection chains. Tool call validation verifies authorized scope before each invocation. The MCP specification evolved rapidly over the course of 2024 and 2025, moving from no authentication in November 2024 to OAuth 2.1 with PKCE in March 2025, Resource Server classification with RFC 8707 Resource Indicators in June 2025, and URL-based client registration with enhanced authorization flows in November 2025~\cite{mcp_spec_2025}. CVE-2025-49596 in MCP Inspector (CVSS 9.4) demonstrated that elementary session tokens and origin verification were initially absent~\cite{oligo2025mcpinspector}. Practitioner guidance has converged on several complementary control categories for MCP deployments, including per-user authentication, provenance tracking, containerized sandboxing, inline policy enforcement, and centralized governance via private registries.

\item[Monitoring and runtime verification.] Agent behavior audit trails logging all tool calls, decisions, and reasoning are essential for forensics and risk management. Anomaly detection must identify tool-call patterns deviating from expected behavior. AgentSpec~\cite{wang2025agentspec} introduces a lightweight DSL for specifying runtime constraints, preventing $>$90\% of unsafe code executions. Pro2Guard~\cite{wang2025pro2guard} extends this with probabilistic model checking via DTMC learning to anticipate unsafe states before violations occur. Human-in-the-loop confirmation remains critical for high-risk operations, yet only 47\% of organizations have implemented GenAI-specific security controls~\cite{microsoft2026cyberpulse}.

\item[Open challenges.] Despite these advances, fundamental gaps persist. Tool poisoning achieves more than 70\% ASR~\cite{wang2025mcptox}. Simon Willison identifies three conditions, namely access to sensitive data, exposure to untrusted content, and the ability to communicate externally, that create an inherent tension most useful agentic applications cannot avoid. OpenAI frames prompt injection defense as a long-term AI security challenge~\cite{openai2025atlas}. Section~\ref{sec:future-agents} details the research agenda these challenges define.
\end{description}

For deployments of agentic systems, the preceding analysis suggests several practical priorities. Tool descriptions and retrieved content should be treated as untrusted inputs, and tool calls should be validated against an authorized, least-privilege scope before execution. Human-in-the-loop confirmation remains advisable for actions with persistent external effects such as file writes, financial transactions, or outbound communication. Where the deployment context permits, structured prompt/data separation can reduce instruction--data confusion. Systems should also retain auditable tool-call traces, approval events, and related decision records sufficient for forensic review and anomaly detection. Finally, MCP servers should not be composed by default across trust boundaries; per-server trust assessment, provenance checks, and sandboxed execution remain important safeguards.


\begin{table}[!htbp]
\centering
\caption{Emerging defenses for agentic AI systems.}
\label{tab:agentic-defense}

\fontsize{7}{9}\selectfont
\setlength{\tabcolsep}{5pt}
\renewcommand{\arraystretch}{1.3}

\begin{tabular}{l l l l l}
\toprule
\textbf{Category} & \textbf{Defense} & \textbf{Mechanism} & \textbf{Target Threat} & \textbf{Status} \\
\midrule

\rowcolor{rowgray}
Architecture
  & Gemini multi-layer~\cite{google2025geminisecurity}
  & Origin sets + alignment critics
  & IPI, tool abuse
  & Production \\

Architecture
  & Instruction hierarchy~\cite{wallace2024instruction}
  & System $>$ user $>$ tool priority
  & Privilege confusion
  & Research \\

\rowcolor{rowgray}
Protocol
  & OAuth 2.1 + PKCE~\cite{mcp_spec_2025}
  & Authentication retrofit
  & Unauthorized access
  & Deployed (MCP) \\

Runtime
  & AgentSpec~\cite{wang2025agentspec}
  & DSL constraint enforcement
  & Unsafe code execution
  & Research \\

\rowcolor{rowgray}
Runtime
  & Pro2Guard~\cite{wang2025pro2guard}
  & Probabilistic model checking
  & Pre-violation detection
  & Research \\

Runtime
  & Progent~\cite{shi2025progent}
  & Programmable privilege control
  & Excessive permissions
  & Research \\

\rowcolor{rowgray}
Governance
  & IMDA framework~\cite{imda2026agenticai}
  & Agent-specific risk tiers
  & Regulatory gap
  & Published \\

\bottomrule
\end{tabular}
\end{table}

\begin{figure}[!htbp]
\centering
\includegraphics[width=0.95\linewidth]{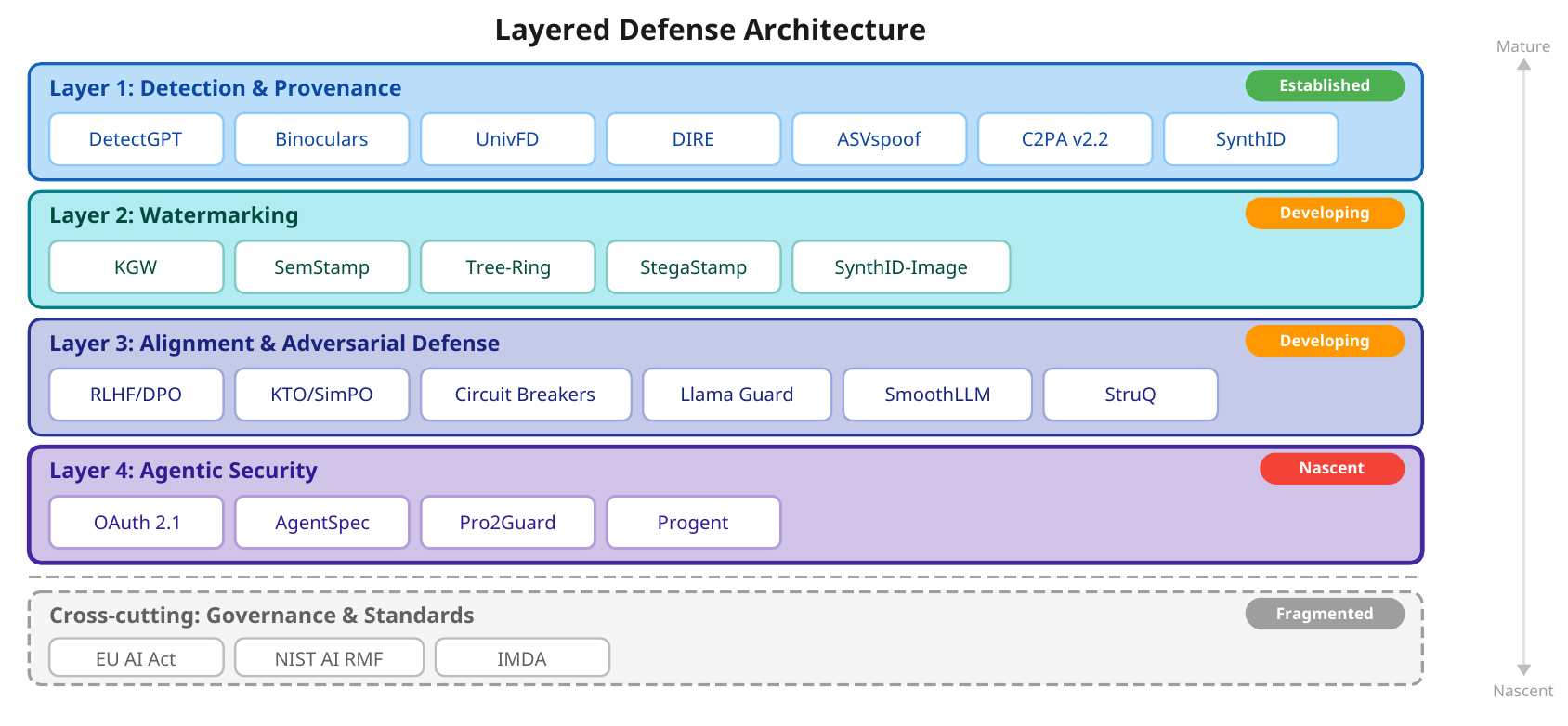}
\caption{Layered defense architecture for AIGC security. Maturity decreases from top to bottom. Maturity levels reflect deployment prevalence and empirical validation: Established = widely deployed with documented effectiveness; Developing = active research with partial deployment; Nascent = primarily academic with limited real-world validation; Fragmented = multiple incompatible approaches without coordination (governance is labeled Fragmented because three distinct paradigms exist but lack mutual recognition).}
\label{fig:defense-layers}
\end{figure}

\section{Discussion and Outlook}
\label{sec:discussion}

The preceding sections show that each major generative AI capability creates security challenges that existing technical defenses and governance arrangements address only partially. This section therefore turns to the main unresolved research problems, the governance gap surrounding agentic AI, and the broader conclusions that follow from the paper.

\subsection{Securing Autonomous AI Agents}
\label{sec:future-agents}

Current defenses for agentic AI address specific attack surfaces, but they provide limited assurance once multiple tools, agents, and trust boundaries interact. This is one of the central difficulties of agentic security, because local protections do not readily compose into system-level guarantees. Recent work on tool poisoning, cross-agent trust exploitation, and the data-access/untrusted-input/external-action tension discussed in Section~\ref{sec:agentic-security} suggests that this composability gap now defines the research frontier. NIST launched a dedicated AI Agent Standards Initiative in February 2026~\cite{nist2026agentstandards}, but no binding standard yet exists. Against this background, we highlight four research questions:

\begin{enumerate}[nosep,leftmargin=1.5em]

    \item \textbf{Formal agent trust models.} Existing agent systems rely on ad hoc trust assumptions across delegation chains, yet a general framework for representing, updating, and revoking trust across agents, tools, and intermediaries remains underdeveloped. A central open problem is whether trust can be revised dynamically under adversarial conditions while still supporting useful delegation.

    \item \textbf{Instruction-data separation.} As discussed in Section~\ref{sec:threat-landscape}, current models do not provide reliable separation between instructions and untrusted data. This leaves an architectural question unresolved: whether alternative designs, such as structured query separation~\cite{chen2025struq} or non-autoregressive architectures, can improve that separation without undermining the instruction-following flexibility that makes these systems useful.

    \item \textbf{Capability-security tension.} Recent work suggests that stronger instruction-following can increase susceptibility to tool poisoning and indirect prompt injection~\cite{wang2025mcptox,yi2025benchmarking} in some settings. That relationship is not universal, however, since more capable models often perform better on standard safety benchmarks such as XSTest~\cite{rottger2024xstest}. The open question is therefore not simply whether stronger models are less secure, but under what conditions capability amplifies vulnerability and whether alignment techniques can be designed to selectively improve robustness to adversarial instructions without degrading compliance with legitimate ones.

    \item \textbf{Composable runtime verification.} Current monitoring
systems, including AgentSpec~\cite{wang2025agentspec} and
Pro2Guard~\cite{wang2025pro2guard}, are mainly designed for single-agent
settings. Moreover, existing approaches adopt incompatible formalisms, for example, StateFlow~\cite{wu2024stateflow} uses finite state machines for workflow sequencing. Extending runtime verification across delegated actions, shared context, and multi-agent boundaries remains difficult,
especially under adaptive adversaries.

The difficulty is not only engineering complexity but a mismatch in
execution assumptions. Emerging agent abstractions such as
AIOS~\cite{mei2024aios} and LangGraph organize execution through
scheduler, syscall, and state-graph metaphors borrowed from
deterministic computing. These abstractions are useful for workflow control, but they do not yield security guarantees. Formal constraint frameworks such as Formal-LLM~\cite{li2024formal}, which use pushdown automata to enforce valid planning, target functional correctness rather than adversarial robustness. This leaves open whether they can ground security verification under hostile input, because LLM-based agents, under stochastic decoding, need not reproduce the
same action sequence from the same high-level input. As a result,
properties verified on one execution trace may not transfer to the
next, and rollback may require constraining sampling rather than
merely restoring inputs.

Existing responses remain partial. Progent~\cite{shi2025progent}
shows that strict privilege control can drive attack success close to
zero, but its access-control model assumes deterministic subject
behavior. Formal-LLM~\cite{li2024formal} and
Pro2Guard~\cite{wang2025pro2guard} instead point toward frameworks that
constrain or verify stochastic execution directly. One possible
extension is probabilistic capability tokens that attenuate
permissions based on estimated alignment confidence, although whether
such mechanisms can be made stable, auditable, and composable across
multi-agent boundaries remains open.

\end{enumerate}

\subsection{Open Problems in Detection, Robustness, and Evaluation}
\label{sec:future-open}

Three additional problems, while less central to this paper's
content-to-action argument, remain important for the wider AIGC
security literature.

Unified multi-modal detection. Section~\ref{sec:detect-multimodal}
showed that detection remains separated by modality and degrades sharply
on in-the-wild content. One reason is that text, image, and audio
detectors rely on different signal types, including token probability
patterns, spatial frequency artifacts, and spectral cues, which do not
yet fit naturally within a single theoretical framework. An open
question is whether foundation-model features can capture
modality-agnostic indicators of synthetic origin, or whether the shared
mathematical structure of diffusion and autoregressive generation can be
used to derive more transferable representations~\cite{ojha2023towards}. Watermarking and privacy. Section~\ref{sec:watermark} identified
tensions among quality, strength, and robustness, together
with a robustness-spoofing tradeoff that limits practical deployment~\cite{pang2024no}.
 Zhang et al.\
derive strong impossibility results for watermarking under natural
assumptions~\cite{zhang2023watermarks}, while generative models can
memorize training data at rates far above ordinary retrieval~\cite{nasr2023scalable}
and differential privacy remains difficult to maintain at frontier
scale~\cite{tramer2022position}. The most important directions are
therefore semantic watermarks that remain robust under paraphrase and
training methods that reduce memorization without incurring the full
cost of differential privacy. Dynamic evaluation. Static benchmarks are easily weakened by
saturation and contamination, and agentic safety evaluation remains at
an early stage. In AgentSafetyBench, no tested agent scores above
60\%~\cite{zhang2024agent}, while MLCommons AILuminate notes
that benchmark success has little predictive value for deployment
safety~\cite{ghosh2025ailuminate}. This leaves two problems unresolved:
how to design evaluation procedures that remain resistant to
contamination, and how to define deployment-facing safety metrics that
are more informative than benchmark pass rates.

\subsection{The Governance Gap}
\label{sec:governance}

The global AIGC governance landscape in March 2026 is organized around three paradigms. The EU model is prescriptive and rights-centric, with the AI Act~\cite{eu_ai_act_2024} imposing risk-tiered obligations and penalties reaching 35M Eurodollar or 7\% of global turnover. The US model combines voluntary federal standards through the NIST AI RMF~\cite{nist_airmf_2023} with a state-level patchwork that includes California's SB~53~\cite{california_sb53_2025} and the Take It Down Act~\cite{us_congress_take_it_down_2025}. The China model is centrally administered and application-specific, with mandatory algorithm filing, corpus safety review, and active enforcement~\cite{china_cac_genai_2023}. International coordination through the AI Safety Summit series and the OECD Principles~\cite{oecd_ai_principles_2024} has produced soft law, but not binding cross-border standards.

In all, these frameworks still provide little direct treatment of agentic AI. Current frontier safety policies, the NIST AI RMF, ISO/IEC 42001, and the EU AI Act contain no references to ``agent,'' ``agentic,'' or autonomous AI systems. Yet AI agents raise risks that are not well captured by content-centered governance alone. Agentic systems act on external systems, access tools dynamically, execute multi-step plans in which errors can cascade, maintain persistent memory that is vulnerable to manipulation, and delegate across agent boundaries in ways that fragment accountability. Existing governance frameworks were not designed with these properties in view.

The deployment timeline provides context for this structural gap. Anthropic launched Computer Use in October 2024, OpenAI released Operator in January 2025, and Claude Code reached general availability in May 2025, reportedly reaching \$1B in annualized revenue by November~\cite{anthropic2025claudecode1b,openai2025codex}. None of these products launched under a binding agent-specific governance standard. A vendor security disclosure by Anthropic reported that Claude Code had been used in what the company described as an AI-orchestrated cyber espionage campaign, with AI executing 80--90\% of operations autonomously~\cite{anthropic2025cyberespionage}. Responses remain early and largely non-binding. Examples include the OWASP Top~10 for Agentic Applications in December 2025~\cite{owasp_agentic_top10_2025}, Singapore's IMDA framework in January 2026~\cite{imda2026agenticai}, and NIST's AI Agent Standards Initiative in February 2026~\cite{nist2026agentstandards}.

This lag matters not only as a timing problem but also as a practical limit on technical defenses. Several of the countermeasures discussed in Section~\ref{sec:countermeasures} depend on forms of institutional coordination that remain weak or absent. The C2PA provenance standard relies on cryptographic metadata that social media platforms routinely strip on upload (Section~\ref{sec:provenance}), a gap that existing regulation leaves largely untouched. Watermark verification depends on cross-provider standardization that remains absent (Section~\ref{sec:wm-text}), with no clear industry body or regulatory mandate driving convergence. The MCP specification evolved from no authentication at launch to OAuth~2.1 within four months (Section~\ref{sec:agentic-security}), but registry governance, tool-description auditing, and server certification still lack a clear institutional locus. When an agent delegates a task through an MCP tool chain and causes harm, existing liability frameworks also provide no settled answer as to where responsibility lies. These examples suggest that the governance gap is not only a matter of delay. It also reflects a structural mismatch, because several technical defenses presuppose forms of coordination that current governance arrangements do not yet provide.
\begin{figure}[!htbp]
\centering
\includegraphics[width=0.95\linewidth]{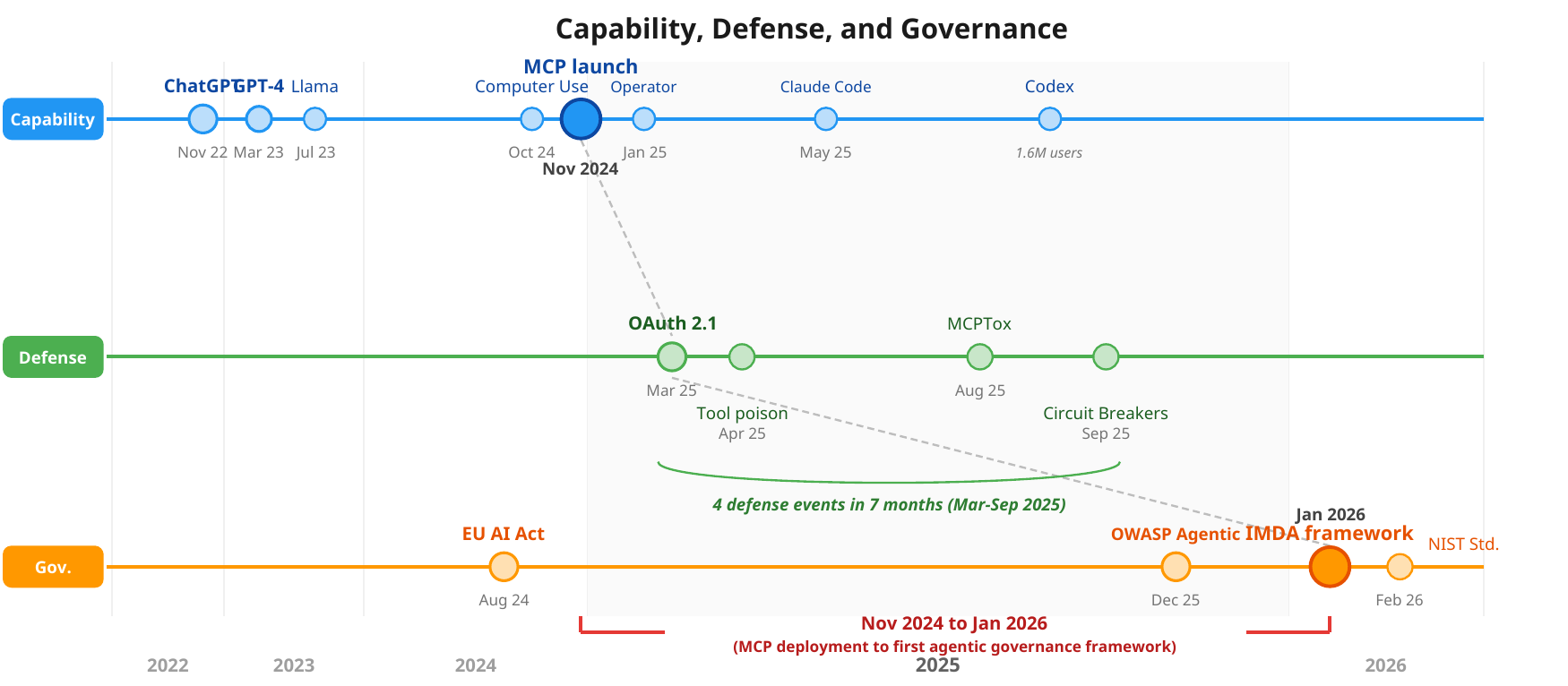}
\caption{Observed ordering across capability deployment, defensive response, and governance frameworks. The 14-month gap from MCP deployment (November 2024) to the first agentic governance standard (January 2026)  illustrates the coordination gap discussed in this section.}
\label{fig:temporal-misalignment}
\end{figure}

\subsection{Limitations}

The analytical framework proposed in this paper has several boundaries that should be stated clearly.

First, the attack surface taxonomy organizes differences in attack complexity and defense difficulty across levels, rather than the actual path taken by every attacker. In practice, an agent system that directly faces users may be exposed to Level~4 tool poisoning attacks from the first day of deployment, without requiring the attacker to pass through Levels~1 to~3. Its value lies in explaining why lower-level defenses fail structurally when facing higher-level attacks, rather than in describing a fixed sequence of escalation.

Second, Levels~1 to~4 are each supported by multiple independent studies and real-world incidents, whereas the current evidence for Level~5 mainly comes from a limited number of early studies. Level~5 in this paper should therefore be understood more as a predictive category based on current trends than as a fully validated threat level.

Third, the recurring ordering pattern is based on observations from three cases, namely SQL injection, about six years, IoT security, about four years, and agentic AI, about fourteen months. The endpoints in these three cases are defined differently, including industry best practice guidance, OWASP lists, and governance frameworks, so they are not sufficient to support a predictive theory. However, the ordering of capability deployment, defense response, and governance development is consistently observed across all three cases, and this ordering is itself a meaningful empirical finding.

Finally, the sources used in this paper span peer-reviewed publications, institutional risk reports, vendor security disclosures, and standards documents. Given the rapid development of agentic AI security, some key arguments rely on sources that have not yet gone through peer review. We have distinguished among different types of evidence, but readers should interpret the reliability of specific claims accordingly.

\section{Conclusion and Outlook}

In this paper, we have shown that the four capability-related properties of AIGC each face a different type of defensive bottleneck. High-fidelity generation faces detector fragility rather than the absence of detectors. Low-cost large-scale production faces a lack of cross-provider coordination in watermarking and provenance mechanisms. Controllability faces the fact that governance frameworks do not yet treat cross-modal identity forgery as a unified category. Autonomous agents face the fact that existing security mechanisms were designed for deterministic systems rather than for LLM-based agents with stochastic execution.

The response mechanisms for these four bottlenecks are not the same. Adversarial threats against AI systems are more likely to trigger technical responses within the same research cycle. Social-level safety harms depend more on platform policy, cross-institutional coordination, and legislation, and their response cycle is longer. Governance lag therefore produces more serious consequences on the safety side than on the security side. The recurring ordering observed across the cases discussed in this paper reinforces this point. Across the cases discussed here, deployment has tended to move faster than defense, and defense faster than governance. Harm at the content level can sometimes still be removed after release, but an agent may already have written files, called APIs, or transferred funds before review can intervene. A governance gap of the same length therefore allows a much larger amount of irreversible harm.

If the response window continues to shrink, the next wave of capability shifts, including large-scale multi-agent orchestration, may leave even less time for standards development and governance construction. At the same time, all levels in the current framework assume that agent input, output, and communication take place in a discrete natural language space, which means that auditing, filtering, and human oversight remain possible at least in principle. Emerging directions in latent reasoning and latent multi-agent communication~\cite{zou2025latent} would replace discrete token-based communication with continuous vector flows, and may fundamentally remove this assumption of auditability. Whether security guarantees can be maintained when inter-agent communication moves beyond the discrete token boundary is among the open questions that will shape the next phase of this field.

\bibliographystyle{cas-model2-names}
\bibliography{reference.bib}
\end{document}